\documentclass{iaus}
\usepackage{graphicx}

\title[Wind of Cyg\,OB2 No.\,12] %% give here short title %%
{The Inhomogeneous Wind of the LBV Candidate Cyg\,OB2 No.12}

\author[Valentina G. Klochkova, Eugene L. Chentsov, \& Anatoly Miroshnichenko]   %% give here short author list %%
{Valentina Klochkova$^1$, Eugene Chentsov$^1$, \and Anatoly
Miroshnichenko$^2$}

\affiliation{$^1$Special Astrophysical Observatory RAS,
         Nizhnij Arkhyz, 369167, Russia \\
email: {\tt valenta@sao.ru, echen@sao.ru} \\[\affilskip]
$^2$Dept. of Physics \& Astronomy, University of North Carolina at Greensboro, \\
P.O. Box 26170, Greensboro, NC 27402--6170, USA  \\email: {\tt
a$\_$mirosh@uncg.edu}}

\pubyear{2010}
\volume{272}  %% insert here IAU Symposium No.
\pagerange{1--2}
 \jname{Active OB stars: structure, evolution,
mass loss and critical limits} \editors{C. Neiner, G. Wade, G.
Meynet \& G. Peters, eds.}
\begin{document}

\maketitle

\begin{abstract} We present the results of high-resolution spectroscopy of
the extremely luminous star Cyg\,OB2~No.\,12. We identified about
200 spectral features in the range 4552--7939\,\AA, including the
interstellar Na\,{\sc I}, K\,{\sc I} lines and numerous very strong
DIBs, along with the He\,{\sc I}, C\,{\sc II}, and Si\,{\sc II}
lines. An MK spectral type we derived for the object is B4.5$\pm$0.5
Ia$^+$. Our analysis of the radial velocity data shows the presence
of a gradient in the stellar atmosphere, caused by both atmospheric
expansion and matter infall onto the star. The H$\alpha$ emission
displays broad Thompson wings, a slightly blue-shifted P\,Cyg type
absorption component and a time-variable core absorption. We
conclude that the wind is variable in time. 
\end{abstract}

\firstsection % if your document starts with a section,
\section{Introduction}

For stars in clusters the evolutionary stage, age, and luminosity
can be determined more reliably, whereas they are rather uncertain
for field stars. It is especially important to study group members
that are rare, such as LBV--stars. From this point of view, young
Cyg\,OB2 association is of special interest. Many unevolved
O/Of--stars have been identified there as well as an LBV candidate
-- the variable star No.\,12. Its luminosity is
log\,L/L$\odot$\,=\,6.26 (\cite[de Jager 1998]{deJager98}) at the
association distance of 1.7 kpc.

\begin{figure}[htb]
% \vspace*{-2.0 cm}
\begin{center}
\begin{tabular}{ll}
\includegraphics[width=2.48in,bb=100 140 540 780,clip]{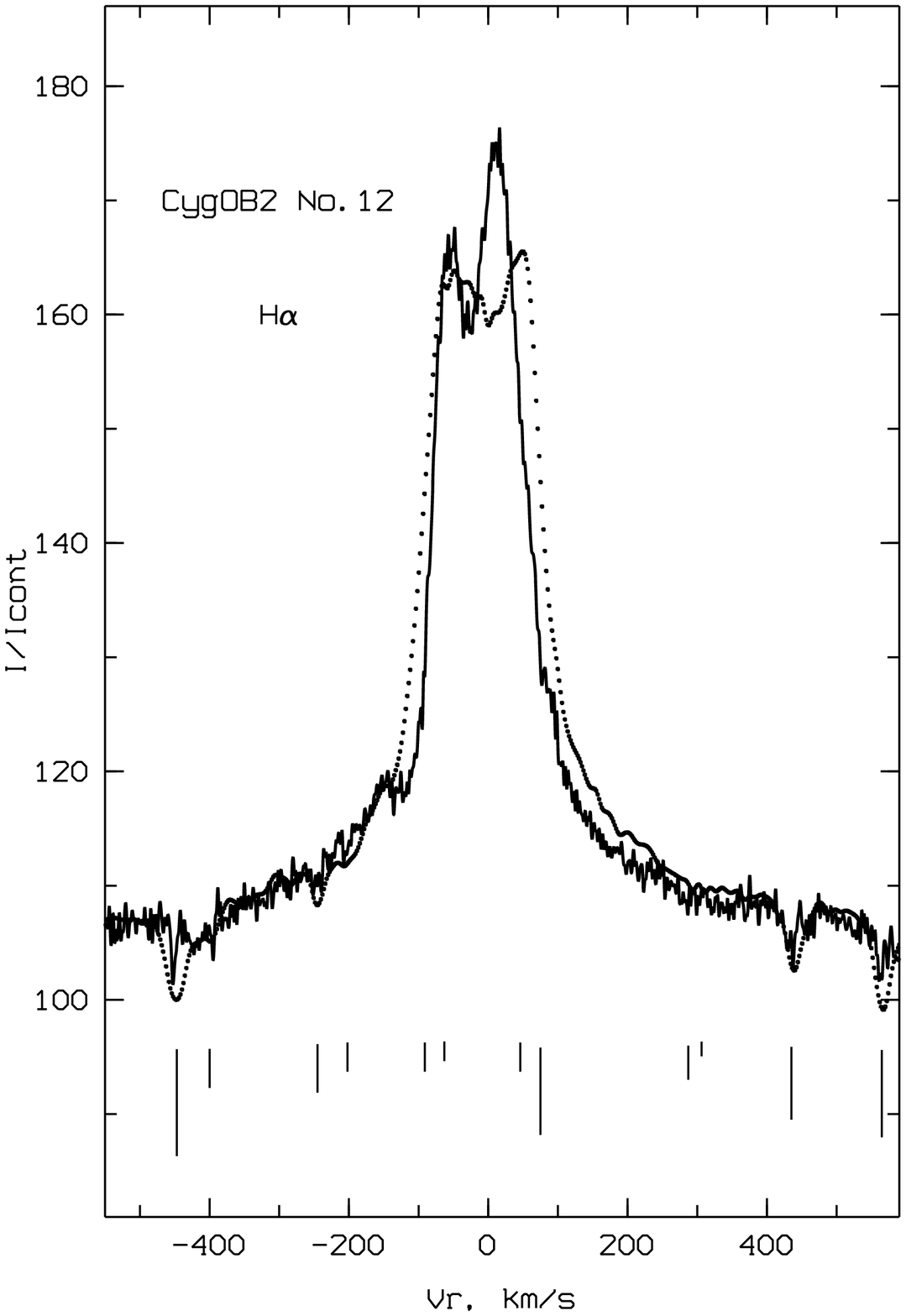} &
\includegraphics[width=2.82in,bb=30 130 540 770,clip]{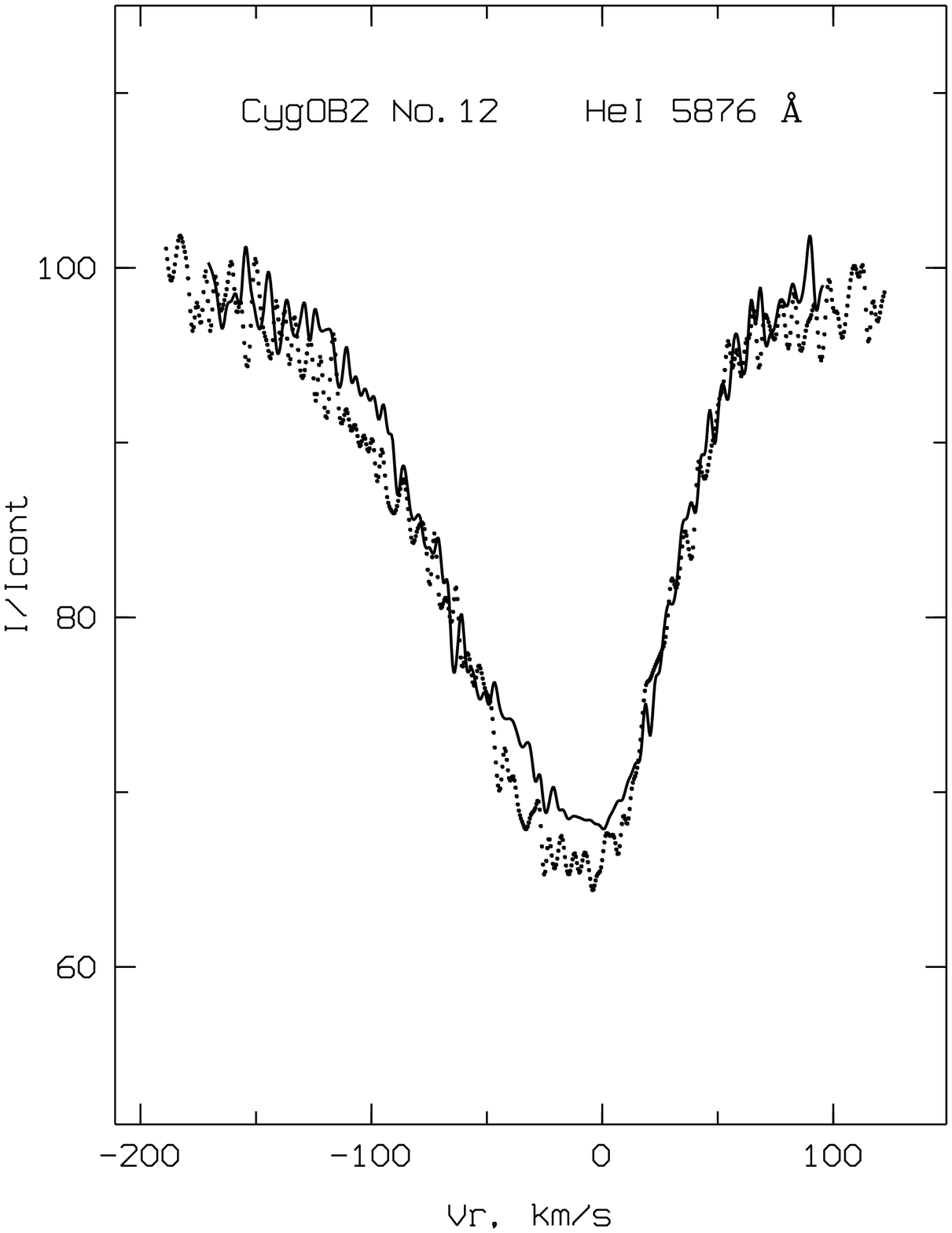} \\
\end{tabular}
% \vspace*{-1.0 cm}
\caption{H$\alpha$ (left) and  He\,{\sc I}\,5876\,\AA\ (right)
profiles in the spectra of Cyg\,OB2~No.\,12 obtained on
June~12,~2001 (dotted line) and April~12,~2003 (solid line). The
absorption features seen within the H$\alpha$ core correspond to
transitions in the line rather than to the telluric spectrum, whose
contribution was carefully removed. The vertical dashes show the
positions of telluric lines, with the dash lengths proportional to
the line strength.} \label{s4-11_fig1}
\end{center}
\end{figure}

\section{Observations and results}

Optical spectra of Cyg\,OB2 No.\,12 were taken using the \'echelle
spectrographs of the 6-meter telescope of the Special Astrophysical
Observatory. On June 12, 2001, we used the PFES spectrograph
(\cite[Panchuk et al. 1998]{Panchuk98}) with a 1040$\times$1170-pixel CCD
at the prime focus and got a spectrum with a resolution of $R
=\lambda/\Delta\lambda \sim$ 15000 (20 km\,s$^{-1}$). Later we used the
NES spectrograph~(\cite[Panchuk et al. 2009]{Panchuk09}) equipped with a
2048$\times$2048-pixel CCD and an image slicer and obtained spectra with
$R \sim$\,60000 (5\,km\,s$^{-1}$) on April 12, 2003 and on
December~8,~2006.

The spectral types we derived for three dates were the same within the
errors: B5.0$\pm$0.5, B4.8$\pm$0.5 and B4.0$\pm$0.5. The
luminosity type is Ia$^+$. The high luminosity is supported by the strong
O\,{\sc I}\,7773\,\AA\ IR--triplet whose equivalent width of 1.14\,\AA\
corresponds to an absolute visual magnitude of M$_{V} <$\,8\,mag.

The radial velocities (V$_{\rm r}$) measured from the absorption line
cores vary with time and with the line intensity. The weakest lines give
V$_{\rm r}$ lower than V$_{\rm sys}$\,=\,$-$11\,km\,s$^{-1}$
(\cite[Klochkova \& Chentsov 2004]{KlochkovaChentsov04}) by 5, 14 and
15\,km\,s$^{-1}$ in 2001, 2003, 2006, respectively, suggesting a variable
expansion rate of the layers where they form. The left panel of
Fig.\,\ref{s4-11_fig1} shows that the H$\alpha$ profile varies with time,
but its principal features are preserved: a strong bell-shaped emission,
with a dip at the short-wavelength slope, a sheared peak, and extended
Thompson wings. The blue-shifted absorption is barely visible in June~2001
and is more pronounced in April~2003, but can be traced at least to
V$_{\rm r}$\,=\,$-$160\,km\,s$^{-1}$ in both cases; i.e., to the same
limit that is reached by the blue wings of the absorption lines of
Si\,{\sc II} and He\,{\sc I} (the latter shown in the right panel of
Fig.\,\ref{s4-11_fig1}).

The wind terminal velocity is $\sim$ 150\,km\,s$^{-1}$. The
intensity inversions in the upper part of the H$\alpha$ profile
indicate that the wind is not uniform. In addition to the high
velocity material mentioned above, it contains a fair amount of
material that is nearly stationary relative to the star or is even
falling onto the stellar surface. Coexistence of lines with direct
and inverse P\,Cygni profiles in the same spectrum, and even
combinations of such features in the profile of the same line leads
us to reject a spherical symmetry wind. It is possible that the slow
part of the wind also contributes to the absorption profiles. So
far, this possibility is supported by the coincident velocities for
the central dips of the H$\alpha$ line and the well-formed cores of
strong absorption lines (He\,{\sc I}\,5876\,\AA\ in 2001 and
Si\,{\sc II}\,6347\,\AA\ in 2003), as well as by the fact that the
blue shift of all the absorption lines in the 2003 spectrum relative
to their positions in 2001 was accompanied by a similar shift of the
central dip in H$\alpha$. At any rate, both the hydrogen lines and
the strongest absorption lines in the visual spectrum
Cyg\,OB2~No.\,12 are partially formed in the wind.

{\bf Acknowledgements}. This research was supported by the Russian
Foundation for Basic Research (project no.~08--02--00072\,a).

\end{document}